\documentclass[aps,twocolumn,showpacs]{revtex4}
\usepackage{amsmath}
\usepackage{epsfig}
\usepackage{epstopdf}
\usepackage{booktabs}
\usepackage{multirow}

\begin{document}

\title{Study of the hidden charm pentaquarks $\Lambda_{c\bar{c}}$ above $4$ GeV }

\author{Xuejie Liu$^1$}\email{1830592517@qq.com}
\author{Hongxia Huang$^1$}\email{hxhuang@njnu.edu.cn(Corresponding author)}
\author{Jialun Ping$^1$}\email{jlping@njnu.edu.cn(Corresponding author)}

\affiliation{$^1$Department of Physics, Nanjing Normal University,
Nanjing 210097, P.R. China}

\begin{abstract}
A dynamical calculation of the strange hidden charm pentaquark systems $\Lambda_{c\bar{c}}$ is performed in the framework of the quark delocalization color screening model. The effective potentials between two clusters are calculated to investigate the interaction between the baryons and mesons. The dynamic calculations indicate that the strange hidden charm pentaquark system with $IJ^{P}=0\frac{1}{2}^{-}$ and $IJ^{P}=0\frac{3}{2}^{-}$ can form bound states with the help of the channel-coupling. The energies of these two system are $4094.3$ MeV and $4207.4$ Mev, respectively. These two $\Lambda_{c\bar{c}}$ states are possible to be intriguing pentaquark candidates which are worth searching in experiments. Whereas, for the systems with isospin $I=1$, the effective potentials of all channels are repulsive, and there is no any bound state for these systems.
\end{abstract}


\maketitle

\setcounter{totalnumber}{5}
\section{\label{sec:introduction}Introduction}
The great age have witnessed tremendous progress on the experimental and theoretical explorations of the multiquark states.
Actually, the possible existence of multiquark states beyond the ordinary hadrons was first proposed by Gell-Mann and Zweig~\cite{1964_1,1964_2}.
As a hot research issue, studying of exotic hadronic states is attractive for both experimentalists and theorists. More and more charmonium-like
and bottomonium-like states and states with open-charm and open-bottom quantum numbers have been reported in experiments as time goes on.
Nowadays, it is an important and interesting topic to look for such states~\cite{exotic}.

In fact, since 2003, experimentalists announced more and more exotic states~\cite{Choi:2003ue,Choi:2007wga,Mizuk:2008me,Mizuk:2009da,Liu:2013dau,Xiao:2013iha,Ablikim:2013mio,Ablikim:2013wzq,Ablikim:2013xfr,Chilikin:2014bkk,Aaij:2014jqa,Ablikim:2015gda,Ablikim:2015swa,Ablikim:2015tbp}, which provides good opportunities to study the nonperturbative color interactions. Some of the states have been considered as good tetraquark candidates~\cite{Swanson:2006st,Zhu:2007wz,Voloshin:2007dx,Drenska:2010kg,Chen:2016qju,Hosaka:2016pey,Richard:2016eis,Lebed:2016hpi,Esposito:2016noz}.
For the pentaquark states, in 2015, the hidden charm pentaquarks (named as $P_{c}$ states) were obtained in the decay $\Lambda_{b}^{0}\rightarrow J/\psi K^{-} p$
by the LHCb Collaborations~\cite{Pc.1,Pc.2,Pc.3}. Then the $P_{c}$ states have attracted much attention and became a hot topic. The main reason is that they are the first exotic baryons discovered experimentally.
Four years later, the LHCb Collaboration updates there results on $P_c$ states, and reported the observation of three new pentaquarks, named as $P_{c}(4312)$,
$P_{c}(4440)$, and $P_{c}(4457)$~\cite{LHCb2}. All of them are observed in the $J/\psi p$ invariant mass spectrum
from the $\Lambda_{b}\rightarrow J/\psi p K^{-}$ decay, corresponding to three peaks in the $J/\psi p$ invariant mass distributions.

To identify the internal structure of $P_{c}$ is the most fundamental problem currently. There are a large number of theoretical studies
about hidden charm pentaquarks in literature, although there is not yet conclusive picture about the structure of $P_{c}$,
including models of hadronic molecules~\cite{Chen:2019bip,Chen:2019asm,Guo:2019fdo,Liu:2019tjn,He:2019ify,Guo:2019kdc}, compact pentaquark
states~\cite{Ali:2019npk,Zhu:2019iwm,Wang:2019got,Giron:2019bcs,Cheng:2019obk,Stancu:2019qga} and hadrocharmonia~\cite{Eides:2019tgv}.
In the framework of quark delocalization color screening model, by considering the baryon-meson scattering process, $P_{c}(4312)$ was proposed
as a hidden-charm molecular state $\Sigma_{c}\bar{D}$ with $J^{P}=\frac{1}{2}^{-}$; $P_c(4380)$ was identified as $\Sigma_c^* D$ state
with $J^{P}=\frac{3}{2}^{-}$, and two resonance states around 4450 MeV were predicted, they are $\Sigma_{c}\bar{D^{*}}$ with $J^{P}=\frac{1}{2}^{-}$
and $J^{P}=\frac{3}{2}^{-}$, respectively~\cite{Pc_huang}. In Ref.~\cite{Pc_R.Chen}, three narrow structures $P_{c}(4312)$, $P_{c}(4440)$ and
$P_{c}(4457)$ were depicted as the molecular $\Sigma_{c}\bar{D}$ with $I=\frac{1}{2},J=\frac{1}{2}$, $\Sigma_{c}\bar{D^{*}}$ with
$I=\frac{1}{2},J=\frac{1}{2}$, and $\Sigma_{c}\bar{D}$ with $I=\frac{1}{2},J=\frac{3}{2}$, respectively.
Meanwhile, some different opinions also appeared, in Ref.~\cite{Pc_4312}, authors inferred that the $P_{c}(4312)$ peak was more likely to be a
virtual state instead. It is clear that, theoretically, different interpretations are often not consistent with each other. Sometimes, searches
for the counterparts of these states may offer new insight into their true nature. The hidden charm pentaquark with strangeness $-1$ is one the
counterparts of $P_{c}$ states, the quark component of the system is $udsc\bar{c}$. We label the system as $\Lambda_{c\bar{c}}$ in this work.

Actually, there are a lot of work of the $P_{c}$ counterparts $\Lambda_{c\bar{c}}$. By using the unitary approach, Ref.~\cite{lamda_cc.1,lamda_cc.1.1}
predicted both the narrow $N^{*}$ resonances and the $\Lambda^{*}$ resonances with hidden charm, where the $N^{*}$ represents the pentaquark composed
of $uudc\bar{c}$ and $\Lambda^{*}$ means the pentaquark composed of $udsc\bar{c}$. The impact of the existence of such states was also discussed in
Refs.~\cite{lamda_cc.2,lamda_cc.3}, which showed that it was possible to observe the states by a careful study of the
$\Xi_{b}^{-}\rightarrow J/\psi K^{-} \Lambda$ decay and the $\Lambda_{b}\rightarrow J/\psi \eta \Lambda$ decay. In Ref.~\cite{lamda_cc.4}, a neat peak
was observed in the $J/\psi \Lambda$ invariant mass distribution by studying the $\Lambda_{b}\rightarrow J/\psi K^{0}\Lambda $ reaction.
Ref.~\cite{lamda_cc.5} investigated the $\Lambda_{c\bar{c}}$ in the quark model, and found that the $\Lambda_{c\bar{c}}$ states below 4400 MeV were
all formed from two colorful clusters, color octet $uds$ and $c\bar{c}$, and the spin and parity could be $\frac{1}{2}^{-}$ and $\frac{3}{2}^{-}$.
Several possible decay process of $udsc\bar{c}$ are discussed, and they found many channels should be suppressed by light flavor SU(3) symmetry or
by the heavy quark symmetry or by both of them. Ref.~\cite{lamda_cc.6} calculated the $\Lambda_{c}\bar{D^{*}_{s}}$, $\Sigma_{c}\bar{D^{*}_{s}}$, $\Sigma_{c^{*}}\bar{D^{*}_{s}}$, $\Xi_{c}\bar{D^{*}}$, $\Xi_{c}^{'}\bar{D^{*}}$, $\Xi_{c}^{*}\bar{D^{*}}$ interactions in one boson exchange model.
The results show that the most promising strange hidden-charm molecular pentaquarks are the $\Xi_{c}^{'}\bar{D^{*}}$ state with $IJ^P=0\frac{1}{2}^{-}$
and the $\Xi_{c}^{*}\bar{D^{*}}$ states with $0\frac{1}{2}^{-}$ and $0\frac{3}{2}^{-}$.
In Ref.~\cite{lamda_cc.7}, they calculated the partial widths of possible decay channels, and the result showed that the $\Lambda_{c\bar{c}}(4213)$
and $\Lambda_{c\bar{c}}(4403)$ were formed as pseudoscalar meson baryon molecules, and $\Lambda_{c\bar{c}}(4370)$ can be considered as a
$\Lambda_{c} D_{s}^{*}$ or $\Xi_{c} \bar{D^{*}}$ molecule state. $\Lambda_{c\bar{c}}(4490)$ is assumed as a $\Xi_{c}^{*} \bar{D}$ bound state.
Within the theoretical uncertainties, all of these studies show that there would be rather stable signals of the hidden charm strange states.

To deal with the same hadronic states in different models is helpful to check the model dependence of the results and to figure out the nature
of these $\Lambda_{c\bar{c}}$ states. In this paper, we systematically study the properties of the hidden charm strange pentaquark using the quark
delocalization color screening model (QDCSM), which was proposed particularly to study the multi-quark system. Our purpose is to investigate whether
a bound state or resonance state exists or not. To be more convincible, the channel coupling effect of the hidden charm strange pentaquark is also included.

This paper is organized as follows. In Sec. II, a brief introduction of QDCSM is given. In Sec. III, we calculate the relevant Hamiltonian elements
and present the corresponding results for the hidden charm strange pentaquark, and some discussion is presented as well. The summary is show in the
last section.

\section{The quark delocalization color screening model (QDCSM)}
The quark delocalization color screening model (QDCSM) is an extension of the native quark cluster model~\cite{QDCSM1} and was developed with aim of
addressing multiquark systems. The detail of QDCSM can be found in the Refs.~\cite{wangFan1,you}.
Here, we just present the salient features of the model. The model Hamiltonian is:
\begin{widetext}
\begin{eqnarray}
H &=& \sum_{i=1}^{5} \left(m_i+\frac{\boldsymbol{p}_i^2}{2m_i}\right)-T_{CM}+\sum_{j>i=1}^5\left[V^C(r_{ij})+V^G(r_{ij})+V^B(r_{ij})++V^{BD}(r_{ij})\right],\\
%\iffalse
V^{G}(r_{ij}) &=& \frac{1}{4}\alpha_s \boldsymbol{\lambda}^{c}_i \cdot\boldsymbol{\lambda}^{c}_j
\left[\frac{1}{r_{ij}}-\frac{\pi}{2}\delta(\boldsymbol{r}_{ij})(\frac{1}{m^2_i}+\frac{1}{m^2_j}
+\frac{4\boldsymbol{\sigma}_i\cdot\boldsymbol{\sigma}_j}{3m_im_j})-\frac{3}{4m_im_jr^3_{ij}}
S_{ij}\right] \label{sala-vG} \\
V^{B}(r_{ij}) & = & V_{\pi}( \boldsymbol{r}_{ij})\sum_{a=1}^3\lambda_{i}^{a}\cdot \lambda
_{j}^{a}+V_{K}(\boldsymbol{r}_{ij})\sum_{a=4}^7\lambda_{i}^{a}\cdot \lambda _{j}^{a}
+V_{\eta}(\boldsymbol{r}_{ij})\left[\left(\lambda _{i}^{8}\cdot
\lambda _{j}^{8}\right)\cos\theta_P-(\lambda _{i}^{0}\cdot
\lambda_{j}^{0}) \sin\theta_P\right] \label{sala-Vchi1}   \\
V^{BD}(r_{ij}) & = & V_{D}(\boldsymbol{r}_{ij})\sum_{a=9}^{12}\lambda_{i}^{a}\cdot \lambda_{j}^{a}+V_{D_{s}}
(\boldsymbol{r}_{ij})\sum_{a=13}^{14}\lambda_{i}^{a}\cdot \lambda_{j}^{a}+V_{\eta_{c}}(\boldsymbol{r}_{ij})\lambda_{i}^{15}\cdot \lambda_{j}^{15}\\
V_{\chi}(\boldsymbol{r}_{ij}) & = & {\frac{g_{ch}^{2}}{{4\pi}}}{\frac{m_{\chi}^{2}}{{\
12m_{i}m_{j}}}}{\frac{\Lambda _{\chi}^{2}}{{\Lambda _{\chi}^{2}-m_{\chi}^{2}}}}
m_{\chi} \left\{(\boldsymbol{\sigma}_{i}\cdot\boldsymbol{\sigma}_{j})
\left[ Y(m_{\chi}\,r_{ij})-{\frac{\Lambda_{\chi}^{3}}{m_{\chi}^{3}}}
Y(\Lambda _{\chi}\,r_{ij})\right] \right.\nonumber \\
&& \left. +\left[H(m_{\chi}r_{ij})-\frac{\Lambda_{\chi}^3}{m_{\chi}^3}
H(\Lambda_{\chi} r_{ij})\right] S_{ij} \right\}, ~~~~~~\chi=\pi, K, \eta, D, D_{s}, \eta_{c}  \\
V^C(r_{ij}) &=&  -a_{c}\boldsymbol{\lambda_{i}\cdot\lambda_{j}}[f(r_{ij})+V_{0}],\\
f(r_{ij}) & = &  \left\{ \begin{array}{ll}r_{ij}^2 &\qquad \mbox{if }i,j\mbox{ occur in the same baryon orbit} \\
\frac{1 - e^{-\mu_{ij} r_{ij}^2} }{\mu_{ij}} & \qquad \mbox{if }i,j\mbox{ occur in different baryon orbits} \\
\end{array} \right. \nonumber \\
& & S_{ij} = \left\{ \frac{(\boldsymbol{\sigma}_i
\cdot\boldsymbol{r}_{ij}) (\boldsymbol{\sigma}_j\cdot
\boldsymbol{r}_{ij})}{r_{ij}^2}-\frac{1}{3}\boldsymbol{\sigma}_i \cdot
\boldsymbol{\sigma}_j\right\},\\
& & H(x) = (1+3/x+3/x^{2})Y(x),~~~~~~
Y(x)=e^{-x}/x. \label{sala-vchi2}
%\fi
\end{eqnarray}
\end{widetext}
Where $S_{ij}$ is quark tensor operator; $Y(x)$ and $H(x)$ are standard Yukawa functions;
$T_{cm}$ is the kinetic energy of the center of mass motion, and $\boldsymbol{\sigma},
\boldsymbol{\lambda}^c, \boldsymbol{\lambda}^a$ are the SU(2) Pauli, SU(3) color, SU(3) flavor
Gell-Mann matrices, respectively. The $\Lambda_{\chi}$ is the chiral symmetry breaking scale, and
$\alpha_{s}$ is the effective scale-dependent running quark-gluon coupling constant~\cite{oge}~,
$\frac{g_{ch}^2}{4\pi}$ is the chiral coupling constant for scalar and pseudoscalar chiral field
coupling, determined from $\pi$-nucleon coupling constant through
\begin{equation}
\frac{g_{ch}^{2}}{4\pi}=\left(\frac{3}{5}\right)^{2} \frac{g_{\pi NN}^{2}}{4\pi} {\frac{m_{u,d}^{2}}{m_{N}^{2}}}
\end{equation}
In the phenomenological confinement potential $V^C$, the color screening parameter $\mu_{ij}$ is
determined by fitting the deuteron properties, $NN$ scattering phase shifts, and $N\Lambda$ and $N\Sigma$
scattering cross sections, respectively, with $\mu_{qq}=0.45, \mu_{qs}=0.19$ and $\mu_{ss}=0.08$,
satisfying the relation $\mu_{qs}^2= \mu_{qq}\mu_{ss}$, where $q$ represents $u$ or $d$. When extending to
the heavy-quark case, there is no experimental date available, so we take it as a adjustable parameter.
In the present work, we take $\mu_{cc}=0.001$ and $\mu_{uc}$ is also obtained by the relation $\mu^{2}=\mu_{uu}\mu_{cc}$.
The other symbols in the above expressions have their usual meanings. All parameters, which are
fixed by fitting the masses of baryons with light flavors and heavy flavors, are taken from our
previous work~\cite{Pc_like}, except for the charm quark mass, which is fixed by fitting to the mass of
the heavy baryons and mesons. The values of those parameters are listed in Table~\ref{parameters}.
The calculated masses of baryons and mesons in comparison with experimental values are shown in Table~\ref{baryons}
\begin{table}[ht]
\caption{\label{biaoge}Model parameters:
$m_{\pi}=0.7$ fm$^{-1}$, $m_{K}=2.51$ fm$^{-1}$,
$m_{\eta}=2.77$ fm$^{-1}$, $m_{D}=9.46$ fm$^{-1}$, $m_{D_{s}}=9.98$ fm$^{-1}$, $m_{\eta_{c}}=15.12$ fm$^{-1}$,
$\Lambda_{\pi}=4.2$ fm$^{-1}$, $\Lambda_{K}=5.2$ fm$^{-1}$,
$\Lambda_{\eta}=5.2$ fm$^{-1}$, $\Lambda_{D}=2.4$ fm$^{-1}$, $\Lambda_{D_{s}}=2.4$ fm$^{-1}$, $\Lambda_{\eta_{c}}=2.4$ fm$^{-1}$,
$g_{ch}^2/(4\pi)$=0.54, $\theta_p$=$-15^{o}$.}
\begin{tabular}{cccccc}
 \hline \hline
~~~$b$ (fm)  &~~~$m_{u}$ (MeV) &~~~$m_{s}$ (MeV) &~~~$m_{c}$ (MeV) \\  \hline
0.518  &313  &573 &1700       \\
~~~~~~$a_{c}$ (MeV fm$^{-2}$) & $V_{0}^{qq}$ (fm$^{2}$)  &~~~$V_{0}^{q\bar{q}}$ (fm$^{2}$)  &~~~$\alpha_{s}^{uu}$   \\
 58.03 &-1.2883 &-0.2012   &0.5652    \\
~~~$\alpha_{s}^{us}$   &~~~$\alpha_{s}^{ss}$ &~~~$\alpha_{s}^{u\bar{u}}$ &$\alpha_{s}^{u\bar{s}}$ \\
0.5239 & 0.4506  & 1.7930 &1.7829  \\
~~~$\alpha_{s}^{s\bar{s}}$   &$\alpha_s^{uc}$ &$\alpha_s^{sc}$  &$\alpha_s^{u\bar{c}}$ \\
 1.5114&0.2031 &0.5675  &1.2859  \\
 ~~~$\alpha_s^{c\bar{c}}$ &$\alpha_s^{s\bar{c}}$\\
 1.6278 &1.5098\\
 \hline\hline
\end{tabular}
\label{parameters}
\end{table}

\begin{table}[ht]
\caption{The Masses (in MeV) of the ground baryons and mesons. Experimental values are taken
from the Particle Data Group (PDG)~\cite{PDG}.}
\begin{tabular}{lcccccccccc}
\hline\hline
&~$J/\psi$~  &~$\eta_{c}$~  &~$\bar{D}$~ &~$\bar{D^{*}}$~ &~$D_{s}$~ &~$D_{s}^{*}$~ \\
\hline
Expt. &3096&  2983&  1869&  2007 &  1968  &2112\\
Model &3005 & 2983 &1913 &2007 &1967 &2027 \\
\hline
& ~$\Lambda$~ & ~$\Lambda_{c}$~  & ~$\Xi_{c}$~ & ~$\Xi_{c^{'}}$~ & ~$\Xi^{*}_{c}$~ & ~$\Sigma_{c}$~ &~$\Sigma^{*}_{c}$~ &~$\Sigma$~  &~$\Sigma^{*}$~\\
\hline
 Expt.& 1115 & 2286 & 2467  & 2577 & 2645 & 2455 & 2520 &1189  &1385\\
 Model & 1123 & 2286 & 2467 & 2538 & 2552 & 2474 & 2485 &1232  &1361\\
\hline
\end{tabular}
\label{baryons}
\end{table}

In QDCSM, the quark delocalization is realized by specifying the single particle orbital
wave function of QDCSM as a linear combination of left and right Gaussian, the single
particle orbital wave functions used in the ordinary quark cluster model,
\begin{eqnarray}
\psi_{r}(\boldsymbol{r},\boldsymbol{s}_{i},\epsilon)&=&(\phi_{R}(\boldsymbol{r},\boldsymbol{s}_{i})
  +\epsilon\phi_{L}(\boldsymbol{r},\boldsymbol{s}_{i}))/N(\epsilon), \label{rl1} \\
\psi_{l}(\boldsymbol{r},\boldsymbol{s}_{i},\epsilon)&=&(\phi_{L}(\boldsymbol{r},\boldsymbol{s}_{i})
  +\epsilon\phi_{R}(\boldsymbol{r},\boldsymbol{s}_{i}))/N(\epsilon), \label{rl2} \\
N(\epsilon)&=& \sqrt{1+\epsilon^2+2\epsilon e^{-s^2_{i}/{4b^2}}},\\
\phi_{R}(\boldsymbol{r},\boldsymbol{s}_{i})&=&(\frac{1}{\pi b^2})^{\frac{3}{4}}
e^{-\frac{1}{2b^2}(\boldsymbol{r}-\frac{2}{5}s_{i})^2},\\
\phi_{L}(\boldsymbol{r},\boldsymbol{s}_{i})&=&(\frac{1}{\pi b^2})^{\frac{3}{4}}
e^{-\frac{1}{2b^2}(\boldsymbol{r}+\frac{3}{5}s_{i})^2},
\end{eqnarray}
The $\boldsymbol{s}_{i}$, $i=1,2,..., n$, are the generating coordinates, which are introduced to
expand the relative motion wave function~\cite{si_1,si_2,si_3}. The mixing parameter
$\epsilon(s_{i})$ is not an adjusted one but determined variationally by the dynamics of the
multi-quark system itself. This assumption allows the multi-quark system to choose its
favorable configuration in the interacting process. It has been used to explain the cross-over
transition between the hadron phase and the quark-gluon plasma phase~\cite{si_4}.
%From the expressions Eq. (\ref{rl1}) and (\ref{rl2}), we can see that the property of
%$\psi_{r}(\boldsymbol{r},\boldsymbol{s}_{i},\epsilon)$ and
%$\psi_{l}(\boldsymbol{r},\boldsymbol{s}_{i},\epsilon)$ under the space inversion is same as that of
%$\phi_{R}(\boldsymbol{r},\boldsymbol{s}_{i})$ and $\phi_{L}(\boldsymbol{r},\boldsymbol{s}_{i})$,
%which is independent of $\epsilon$. So the parity of the system with delocalized single-particle
%wave functions is the same as that of system with Gaussian as the single-particle wave functions,
%$P=(-1)^{L+1}$ with $L$ is the orbital angular momentum between two subclusters if the subclusters are
%in the ground states. In fact, F. Stancu and L. Wilets showed this property in their paper
%(Fig.~2)~\cite{Stancu}.

\section{The results and discussions}
In present work, we systematically investigate the low-lying $S$-wave hidden charm strange pentaquark $\Lambda_{c\bar{c}}$ systems.
We consider the states with the baryon-meson molecular structure, and the parity of the states is negative. The orbital angular
momentum $L$ between two clusters is set to $0$, and the total angular momentum and parity can be $J^{P}=\frac{1}{2}^{-}, \frac{3}{2}^{-},
\frac{5}{2}^{-}$. The isospin of the pentaquark states can be $0$ and $1$. All possible channels are listed in Table ~\ref{channels}.

\begin{center}
\begin{table}[h]
\caption{All possible channels for $udsc\bar{c}$ systems.}
\begin{tabular}{lccccccccccc}
\hline\hline
$IJ=0\frac{1}{2}$~~  &$\Lambda_{c}D_{s}$  &$\Lambda_{c}D_{s}^{*}$ &$\Xi_{c}\bar{D}$
                        &$\Xi_{c^{'}}\bar{D}$ &$\Xi_{c}\bar{D^{*}}$  &$\Xi_{c}^{'}\bar{D^{*}}$
                         &$\Xi_{c}^{*}\bar{D^{*}}$   \\
      &                &$\Lambda\eta_{c}$  &$\Lambda J/\psi$\\

$IJ=0\frac{3}{2}$   &$\Lambda_{c}D_{s}^{*}$
                        &$\Xi_{c}\bar{D^{*}}$  &$\Xi_{c}^{'}\bar{D^{*}}$
                         &$\Xi_{c}^{*}\bar{D^{*}}$  &$\Xi_{c}^{*}\bar{D}$  &$\Lambda J/\psi$\\
 $IJ=0\frac{5}{2}$   &$\Xi_{c}^{*}\bar{D^{*}}$   \\

$IJ=1\frac{1}{2}$   &$\Xi_{c}\bar{D}$
                         &$\Xi_{c}^{'}\bar{D}$ &$\Xi_{c}\bar{D^{*}}$  &$\Xi_{c}^{'}\bar{D^{*}}$
                         &$\Xi_{c}^{*}\bar{D^{*}}$  &$\Sigma\eta_{c}$ &$\Sigma J/\psi$\\
               &           &$\Sigma_{c}D_{s}$ &$\Sigma_{c}D^{*}_{s}$ &$\Sigma^{*} J/\psi$ &$\Sigma_{c}^{*}D_{s}^{*}$\\

$IJ=1\frac{3}{2}$  &$\Xi_{c}\bar{D^{*}}$  &$\Xi_{c}^{'}\bar{D^{*}}$
                         &$\Xi_{c}^{*}\bar{D^{*}}$  &$\Xi_{c}^{*}\bar{D}$ &$\Sigma^{*}\eta_{c}$ & $\Sigma J/\psi$ &$\Sigma^{*}_{c}D^{*}_{s}$ \\
               &           &$\Sigma_{c}\eta_{c}$ &$\Sigma^{*} J/\psi$ &$\Sigma^{*}_{c} J/\psi$ \\

$IJ=1\frac{5}{2}$  &$\Xi_{c}^{*}\bar{D^{*}}$   &$\Sigma J/\psi$   &$\Sigma_{c}\bar{D^{*}_{s}}$   \\
\hline\hline
\end{tabular}
\label{channels}
\end{table}
\end{center}
\subsection{The effective potential calculation}

\begin{figure*}
\centering
\epsfxsize=7.0in \epsfbox{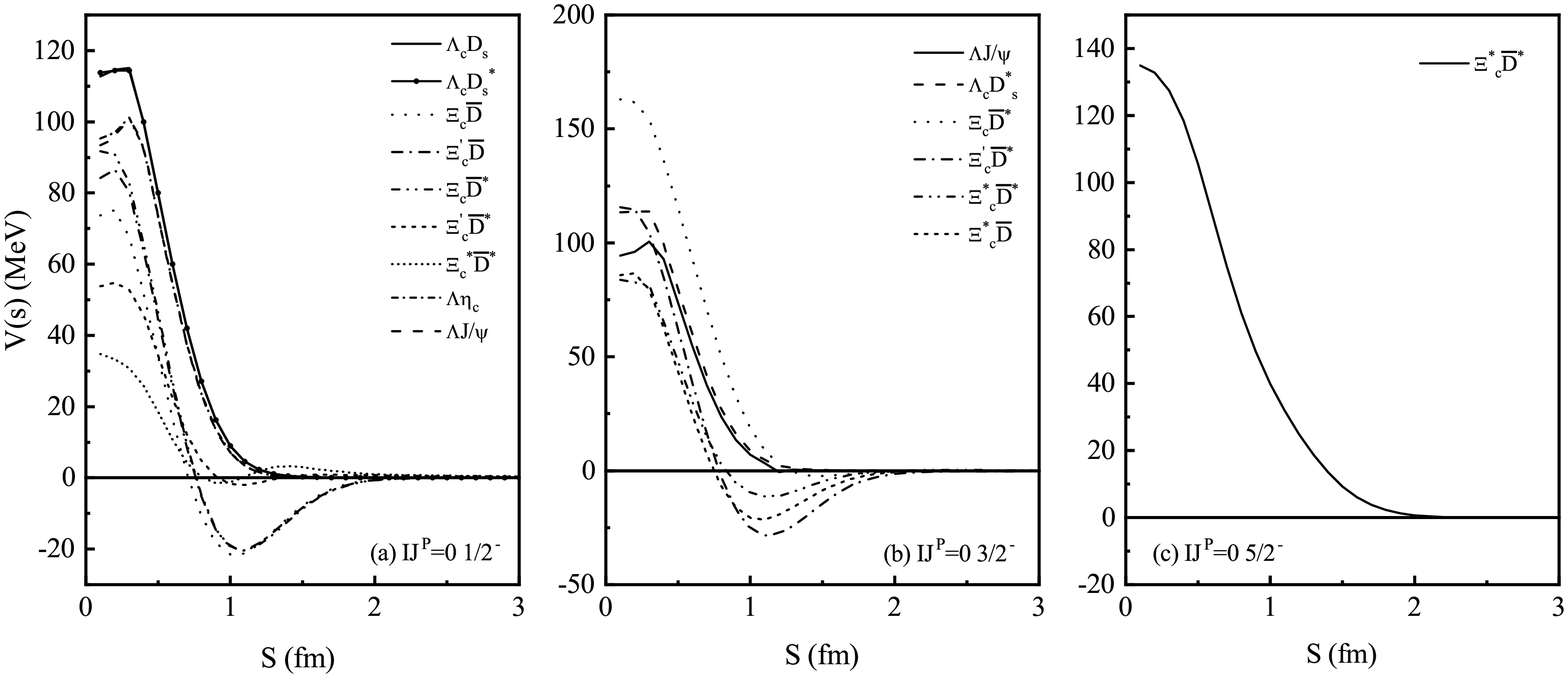} \vspace{-1.05in}
\caption{The effective potentials of every channel in the $\Lambda_{c\bar{c}}$ system with $I=0$.}
\end{figure*}

\begin{figure*}
\epsfxsize=7.0in \epsfbox{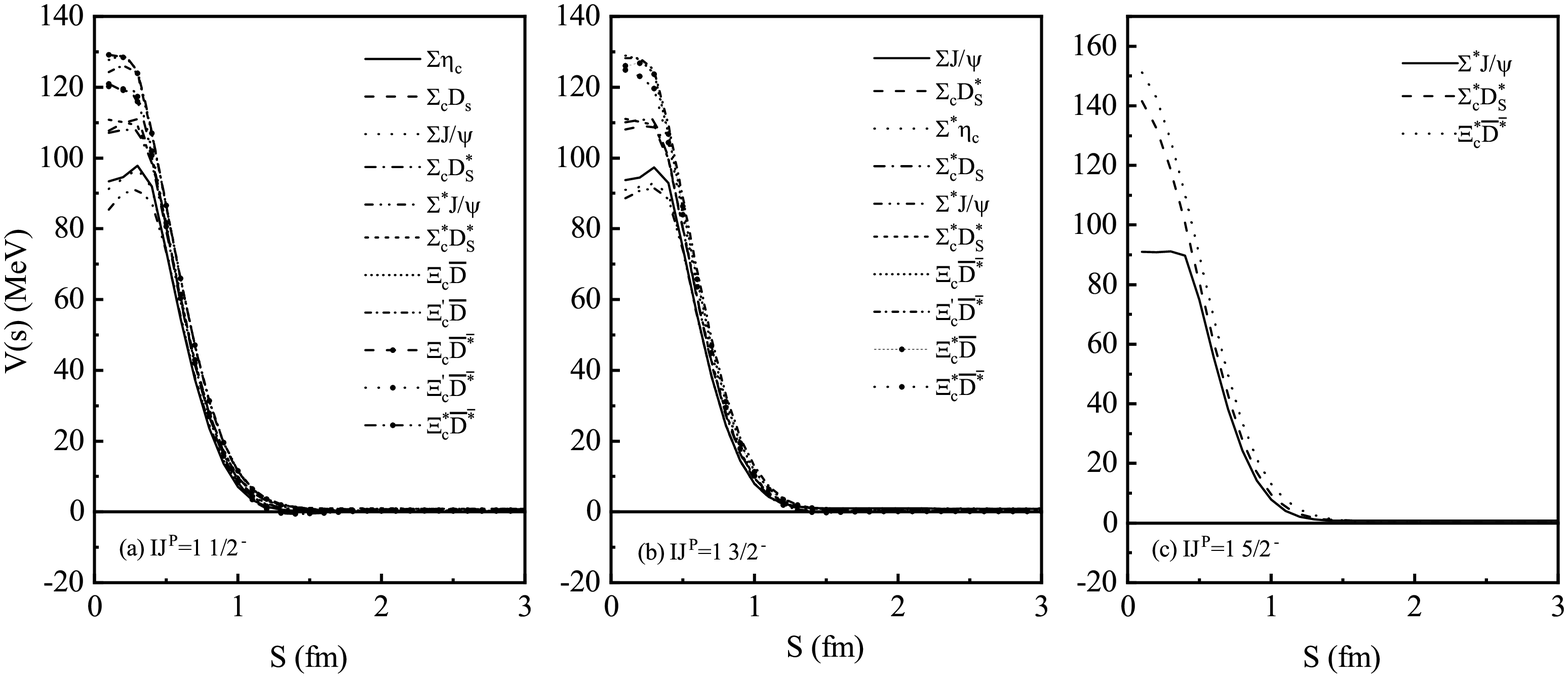} \vspace{-1.05in}
\caption{The effective potentials of every channel in the $\Lambda_{c\bar{c}}$ system with $I=1$.}
\end{figure*}

Since an attractive potential is necessary for forming bound states or resonance states. As the first step, the effective potentials
of all the channels are calculated. The effective potential between two colorless clusters is defined as $V(s)=E(s)-E(\infty)$, where
$E(s)$ is the diagonal matrix element of the Hamiltonian of the system at given generating coordinate $s$. The effective potentials
of the $S$-wave $\Lambda_{c\bar{c}}$ systems with $I=0,1$ are shown in Fig.1 and 2 respectively.

For the $IJ^{P}=0\frac{1}{2}^{-}$ system, the effective potentials of nine channels are shown in Fig.1(a). One can see that the potentials
are attractive for the channels $\Xi_{c}\bar{D}$, $\Xi_{c}^{'}\bar{D}$, $\Xi_{c}\bar{D^{*}}$, $\Xi_{c}^{'}\bar{D^{*}}$,
$\Xi_{c}^{*}\bar{D^{*}}$, $\Lambda\eta_{c}$ and $\Lambda J/\phi$. For the channels $\Lambda_{c}D_{s}$ and $\Lambda_{c}D_{s}$, the potentials
are repulsive, so no bound states or resonances can be formed in these two channels. However, for other channels, the bound states or resonances
are possible due to the attractive nature of the interaction between two hadrons. The attraction between $\Xi_{c}$ and $\bar{D}$ is largest one,
followed by that of the $\Xi_{c}^{'}\bar{D}$ and $\Xi_{c}\bar{D^{*}}$ channel. The $\Xi_{c}^{'}\bar{D^{*}}$ and $\Xi_{c}^{*}\bar{D^{*}}$ have
very weak attractions comparing with $\Xi_{c}\bar{D}$, $\Xi_{c}^{'}\bar{D}$, $\Xi_{c}\bar{D^{*}}$. In addition, the attraction of $\Lambda\eta_{c}$
is almost the same as that of $\Lambda J/\phi$, which is the smallest one during these channels. To confirm whether these states are bound states
or not, a dynamic calculation is needed.

For the the $IJ^{P}=0\frac{3}{2}^{-}$ system (see Fig.1(b)), the potential of the $\Lambda J/\phi$, $\Xi_{c}^{'}\bar{D^{*}}$,
$\Xi_{c}^{*}\bar{D^{*}}$ and $\Xi_{c}^{*}\bar{D}$ channels show an attractive property, while the potentials of other channels are repulsive.
The attraction of the $\Xi_{c}^{'}\bar{D^{*}}$ channel is the strongest, which indicates that the $\Xi_{c}^{'}\bar{D^{*}}$ is more possible to be
a bound state. The attractions of both the $\Xi_{c}^{*}\bar{D^{*}}$ and the $\Xi_{c}^{*}\bar{D}$ channels are smaller than that of
$\Xi_{c}^{'}\bar{D^{*}}$, but it is interesting to explore the possibility of the bound or resonance states of those two channels. For the
$\Lambda J/\phi$, although the attraction of the potential is very small, a bound state is also possible.

For the $IJ^{P}=0\frac{5}{2}^{-}$ system, there is only one channel which is the $\Xi_{c}^{*}\bar{D^{*}}$. Obviously the potential of the
$\Xi_{c}^{*}\bar{D^{*}}$ is repulsive, which means that it is impossible for the $\Xi_{c}^{*}\bar{D^{*}}$ to form a bound state.

For the $IJ^{P}=1\frac{1}{2}^{-}$, $1\frac{3}{2}^{-}$, and $1\frac{5}{2}^{-}$ systems, the potentials of all channels show the repulsive
property, so it is difficult for the systems with $I=1$ to form any bound state.

\subsection{The bound-state calculation}
In order to check whether or not there is any bound state, a dynamical calculation is needed. The resonating group method (RGM)~\cite{RGM1},
a well established method for studying a bound state problem or a scattering one, is employed here. After expanding the relative motion wave
function between two clusters by Gaussians, the integro-differential equation of RGM can be reduced to an algebraic equation, which is the
generalized eigen-equation. The energy of the system can be obtained by solving the eigen-equation. Besides, to keep the matrix dimension manageably
small, the baryon-meson separation is taken to be less than $6$ fm in the calculation. The details of RGM can be found in the Refs.~\cite{RGM1,RGM2}.
The binding energies of every single channel and those with channel coupling are listed in Table ~\ref{bound}, where $E_{c.c}$ stands for the result
of channel-coupling calculation, and $ub$ means that the state is unbound. The symbol of ``$\cdots$" represents that the corresponding channel is
not exist in the system.
Before discussing the features of the states, we should mention how we obtain the mass of these states. Because the quark model cannot reproduce
the experimental masses of all baryons and mesons, the theoretical threshold and the experimental threshold for a given channel is different
(the threshold is the sum of the masses of the baryon and the meson in the given channel). So first, we define the binding energy by
$B=M^{the}-M_{B}^{the}-M_{M}^{the}$, where $M^{the}$, $M_{B}^{the}$, and $M_{M}^{the}$ stand for the theoretical mass of the molecular states,
a baryon, and a meson, respectively. To minimize the theoretical errors, the mass of a bound state can be defined as $M=B+M_{B}^{exp}+M_{M}^{exp}$,
where the experimental values of a baryon and a meson are used.
\begin{table}[ht]
\caption{The binding energies of the $udsc\bar{c}$ systems.}
\begin{tabular}{lcccccccccc}
\hline\hline
Channel &
$IJ^{P}$=$0\frac{1}{2}^{-}$ &$IJ^{P}$=$0\frac{3}{2}^{-}$  &$IJ^{P}$=$0\frac{5}{2}^{-}$ \\\hline
$\Lambda_{c}D_{s}$     &ub   &$\cdots$   &$\cdots$    \\
$\Lambda_{c}D_{s}^{*}$ &ub   &ub         &$\cdots$    \\
$\Xi_{c}\bar{D}$       &ub   &$\cdots$   &$\cdots$    \\
$\Xi_{c}^{'}\bar{D}$   &ub   &$\cdots$   &$\cdots$    \\
$\Xi_{c}\bar{D^{*}}$   &ub   &ub         &$\cdots$     \\
$\Xi_{c}^{'}\bar{D^{*}}$ &ub &ub         &$\cdots$     \\
$\Xi_{c}^{*}\bar{D^{*}}$ &ub &ub         &ub           \\
$\Xi_{c}^{*}\bar{D}$  &$\cdots$  &ub     &$\cdots$     \\
$\Lambda\eta_{c}$     &ub        &$\cdots$ &$\cdots$   \\
$\Lambda J/\psi$      &ub        &ub       &$\cdots$    \\
$B_{c.c}$             &$-3.7$ MeV        &$-3.6$ MeV       &$\cdots$ \\
$E_{c.c}$             &~~4094.3 MeV~~        &~~4207.4 MeV~~       &$\cdots$ \\
\hline\hline
\end{tabular}
\label{bound}
\end{table}

For the $I=0,J^{P}=\frac{1}{2}^{-}$ system, the single-channel calculation shows that none of the single channel is bound, which means that
the attractive interaction of these channels do not leads to the energy of those channels below their corresponding threshold, respectively.
For example, for the $\Xi_{c}\bar{D}$ channels, the interaction between $\Xi_{c}$ and $\bar{D}$ is attractive as shown in Fig. 1(a), but it is
not large enough to form a bound state. At the same time, we also do a channel-coupling calculation, the results of which are shown in Table
\ref{bound} too. A stable state is obtained, the mass of which is lower then the threshold of the lowest channel $\Lambda \eta_{c}$.
This indicates that the $\Lambda_{c\bar{c}}$ system with $I=0,J^{P}=\frac{1}{2}^{-}$ is bound by channel coupling in our quark-model calculation,
and the energy is $4094.3$ MeV. This result shows that the channel coupling effect has an important impact on the existence of the bound state.

For the $I=0,J^{P}=\frac{3}{2}^{-}$ system, although the attraction of the $\Xi_{c}^{'}\bar{D^{*}}$ channel is the largest one, the attraction
is still not large enough to form a bound state. The single channel calculation of other channels shows that none of them can form any bound state
due to the weak attraction. The channel coupling is also considered, which can make the $I=0$, $J^{P}=\frac{3}{2}^{-}$ system to be bound with
a binding energy of $-3.6$ MeV. By using the subtraction procedure we finally obtain the mass of this system, which is $4207$ MeV. This result is
consistent with the prediction in Ref.~\cite{lamda_cc.1.1}, where the $\Lambda_{c\bar{c}}$ state with a mass around $4213$ MeV.

For the $I=0,J^{P}=\frac{5}{2}^{-}$ system, there is only one channel: $\Xi_{c}^{*}\bar{D^{*}}$. As mentioned in last section that the interaction
between $\Xi_{c}^{*}$ and $\bar{D^{*}}$ is repulsive, so it cannot be a bound state here.

For the $IJ^{P}=1\frac{1}{2}^{-}$, $1\frac{3}{2}^{-}$, and $1\frac{5}{2}^{-}$ systems, none of the system is bound whether it is single or
multi-channel coupling computation because of the repulsive interaction.

\begin{figure}[t]
\centering
\epsfxsize=3.5in \epsfbox{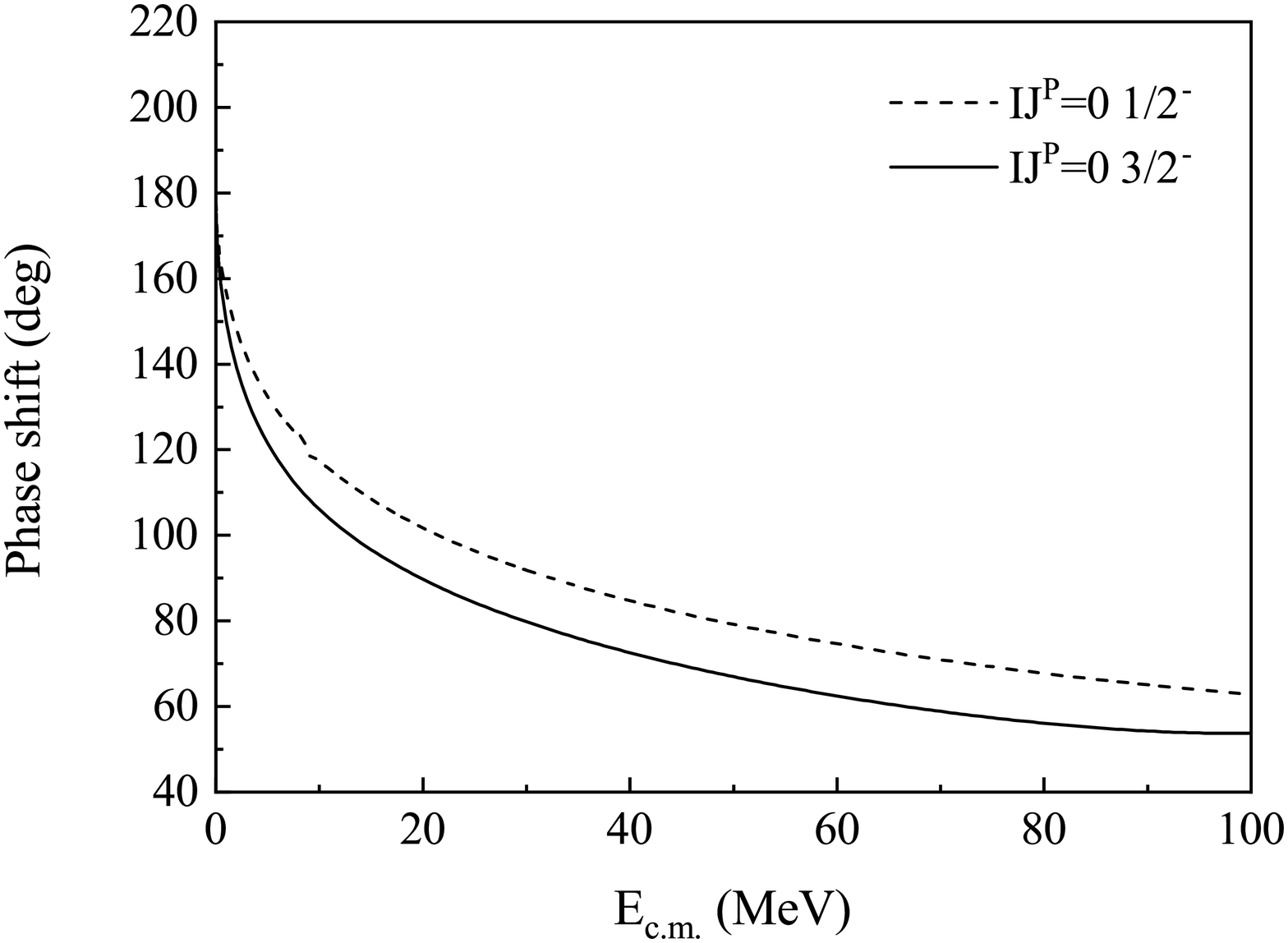} \vspace{-0.05in}
\caption{The phase shifts of the $IJ^{P}=0\frac{1}{2}^{-} \Lambda\eta_{c}$ state and $IJ^{P}=0\frac{3}{2}^{-} \Lambda J/\psi$ state}
\end{figure}
\begin{table}[h]
\caption{Distance between two clusters of $udsc\bar{c}$ systems.}
\begin{tabular}{lc}
\hline\hline
  Channel & Separation (fm) \\
\hline
$IJ^{P}=0\frac{1}{2}^{-}$ &  1.35  \\
$IJ^{P}=0\frac{3}{2}^{-}$ &  1.45  \\
\hline\hline
\end{tabular}
\label{R2}
\end{table}

To check the boundness of the states, we also calculate the low-energy scattering phase shifts of $\Lambda\eta_{c}$ and $\Lambda J/\psi$
with the help of well-developed Kohn-Hulthen-Kato variational method~\cite{KHK}. Fig. 3 illustrates the scattering phase shifts of the
$\Lambda \eta_c$ with $IJ=0\frac{1}{2}^{-}$ and $\Lambda J/\psi$ with $IJ=0\frac{3}{2}^{-}$. It is obvious that in QDCSM, the scattering
phase shifts go to $180^{0}$ at $E_{c.m}\sim 0$ and rapidly decreases as $E_{c.m}$ increases, which implies the existence of a bound state.
The results are consistent with the bound state calculation shown above.

In order to analyse the spacial structure of the $udsc\bar{c}$ pentaquark, we calculated the distances between two sub-clusters of the
$udsc\bar{c}$ systems. From the Table~\ref{R2}, the distances between the two sub-clusters for the two states mentioned above
in the channel-coupling calculations, are around 1.4 fm, which is larger than the size of two sub-clusters. This phenomenon shows that
two sub-clusters tend to stay away from each other, so the structures of the $udsc\bar{c}$ pentaquark is molecular state.

\section{Summary}
Searching for exotic hadronic states is a research field full of challenges and opportunities. With recent experimental progress, more and more
novel phenomena have been revealed in experiments, which has stimulated theorists' extensive interest in studying exotic states. The observation
of $P_c$ states at LHCb inspires many new investigations of the hidden charm pentaquarks. In this situation, it is also interesting to investigate
the partners of $P_{c}$ states. Thus, in the framework of the QDCSM, we perform a theoretical study of the pentaquark systems with quark contents
of $udsc\bar{c}$ ($\Lambda_{c\bar{c}}$) with the aim of looking for any strange hidden charm pentaquarks.

The dynamic calculations show that although all channels with $IJ^{P}=0\frac{1}{2}^{-}$ is unbound in the single channel calculation, a bound state
can be obtained by the effect of channel coupling, and the energy of this system is $4094.3$ MeV.
For the $IJ^{P}=0\frac{3}{2}^{-}$ system, similar results to the case of the $IJ^{P}=0\frac{1}{2}^{-}$ system are obtained. There also exists a
bound state by multi-channel coupling, with the mass of this system is $4207.4$ MeV, which is consistent with the $\Lambda_{c\bar{c}}(4213)$ of
the Ref.~\cite{lamda_cc.1.1}. In addition, these results also indicate that the channel coupling effect in this system has an important influence
on existence of the bound state. The low-energy scattering phase shifts of $\Lambda\eta_{c}$ with $IJ^{P}=0\frac{1}{2}^{-}$ and 
$\Lambda J/\psi$ with $IJ^{P}=0\frac{3}{2}^{-}$ confirm the existence of the bound states. For the system with $IJ^{P}=1\frac{1}{2}^{-}$, 
$1\frac{3}{2}^{-}$, and $1\frac{5}{2}^{-}$, since the interaction of all channels is repulsive, there is no any bound state.

In summary, the observed $P_{c}$ states form LHCb have opened fascinating new avenues of research. In the present work, the result show that
the strange hidden charm pentaquarks $\Lambda_{c\bar{c}}$ with $IJ^{P}=0\frac{1}{2}^{-}$ and $IJ^{P}=0\frac{3}{2}^{-}$ are also intriguing
candidates which should be searched in experiments studies. This is an interesting subject for experiments at high energy accelerator facilities.

\acknowledgments{This work is supported partly by the National Science Foundation of China under
Contract Nos. 11675080, 11175088 and 11535005.}

\end{document}